# xLED: Covert Data Exfiltration from Air-Gapped Networks via Router LEDs

Mordechai Guri, Boris Zadov, Andrey Daidakulov, Yuval Elovici
Ben-Gurion University of the Negev
Cyber Security Research Center
gurim@post.bgu.ac.il; borisza@gmail.com; daidakul@post.bgu.ac.il; elovici@bgu.ac.il

Video: https://www.youtube.com/watch?v=mSNt4h7EDKo
Air-gap research page: http://cyber.bgu.ac.il/advanced-cyber/airgap

*Abstract*— In this paper we show how attackers can covertly leak data (e.g., encryption keys, passwords and files) from highly secure or air-gapped networks via the row of status LEDs that exists in networking equipment such as LAN switches and routers. Although it is known that some network equipment emanates optical signals correlated with the information being processed by the device ('side-channel'), intentionally controlling the status LEDs to carry *any* type of data ('covert-channel') has never studied before. A malicious code is executed on the LAN switch or router, allowing full control of the status LEDs. Sensitive data can be encoded and modulated over the blinking of the LEDs. The generated signals can then be recorded by various types of remote cameras and optical sensors. We provide the technical background on the internal architecture of switches and routers (at both the hardware and software level) which enables this type of attack. We also present amplitude and frequency based modulation and encoding schemas, along with a simple transmission protocol. We implement a prototype of an exfiltration malware and discuss its design and implementation. We evaluate this method with a few routers and different types of LEDs. In addition, we tested various receivers including remote cameras, security cameras, smartphone cameras, and optical sensors, and also discuss different detection and prevention countermeasures. Our experiment shows that sensitive data can be covertly leaked via the status LEDs of switches and routers at a bit rates of 10 bit/sec to more than 1Kbit/sec per LED.

***Keywords— ezfiltration; air-gap; network; optical; covert-channel (key words)***

## I. Introduction

The *air-gap* in network security refers to a situation where the computer network is physically separated from other networks. In particular, air-gapped network are carefully kept isolated form less secure and public networks such as the Internet. Today, air-gapped networks are widely used in military defense systems, critical infrastructure, the financial sector, and other industries [1] [2]. The air-gap isolation is maintained by enforcing strict policies in the organization. These policies include forbidding external unsecured devices and media from connecting to the network and using advanced intrusion detection and prevention systems to eliminate intentional and accidental security breaches. A publicly known example of an air-gapped network is the Joint Worldwide Intelligence Communications System (JWICS), a top secret classified network (Top Secret/SCI) belonging to the United States' Defense Intelligence Agency [3]. JWICS is used to transmit classified documents between the Department of Defense, Department of State, Department of Homeland Security, and Department of Justice.

### A. *Breaching air-gapped networks*

In the past decade it has been shown than even air-gapped networks are not immune to breaches. In order to penetrate such networks, attackers have used complex attack vectors, such as supply chain attacks, malicious insiders, and social engineering [4]. Using these methods, attackers can penetrate an air-gapped network, while bypassing defense measures including firewalls, antivirus programs, intrusion detection and prevention systems (IDS/IPS), and the like. In 2008, a classified network of the United States military was compromised by a computer worm named Agent.Btz [4]. According to the reports [5], a foreign intelligence agency supplied infected thumb drives to retail kiosks near NATO headquarters in Kabul. The malicious thumb drive was put into a USB port of a laptop computer that was attached to United States Central Command. The worm spread further to both classified and unclassified networks. Stuxnet [6] is another well publicized example of an air-gap breach, and other attacks have also been reported [7] [8] [6] [9] [10].

### B. *Leaking data through an air-gap*

After obtain a foothold in an air-gapped network, the attacker may, at some point, wish to retrieve information from the compromised network. For example, an attacker may want to leak encryption keys, keylogging information, or specific files, behaviors commonly used in espionage APTs (Advanced Persistent Threats). While *infiltrating* air-gapped systems has been shown feasible, the *exfiltration* of data from systems without Internet connectivity is a challenging task. Over the years, various out-of-band communication channels have been proposed. These communication channels allow attackers to leak data from network-less computers. When such communication is also covert, it is typically referred as air-gap covert channels. Air-gap covert channels use various types of radiation emitted from computer hardware components. Exploiting electromagnetic radiation is a type of covert channel that has been studied for at least twenty years. In this method, a malware controls the electromagnetic emission from computer parts, such as LCD screens, communication cables, computer buses, and other hardware peripherals [11] [12] [13] [14] [15]. Leaking data over audible and inaudible sounds [16] [17] and via heat emission [18] have also been studied in recent years.

Leaking data via the hard drive activity LED [19], keyboard LEDs [20], and screen power LEDs [21] was also proposed. In these optical methods, data is modulated over the blinks of LEDs in such a way that it can be received optically by remote cameras. Notably, most of the methods are not considered completely covert, since they can easily be detected by people who notice the anomalous LED blinking.

### C. *Our contribution*

In this paper, we examine the threat of leaking data from air-gapped networks via the row of LEDs that exists in network equipment such as switches and routers[1]. The contributions of our research are threefold.

**Novel technical discussion.** The concept of leaking data by controlling the network equipment LEDs has never been studied before. Although it is known that network equipment are emanating optical signals correlated with the information being processed by the device [20], intentionally controlling the status LEDs to modulate and carry data has never studied before. We introduce two types of attacks: firmware level attacks in which malware is installed within the firmware of a network switch or router, and software level attacks in which the malware controls the LEDs from a compromised computer within the network.

**Increased Bandwidth.** Our measurements show that the by using the router status LEDs, malware can exfiltrate data at a rate of more than 1000 bit/sec per LED. We show that the bandwidth can be increased further when multiple LEDs are used. This rate allows the exfiltration of files, keylogging data, and encryption keys relatively quickly.

**The Successful Use of Optical Sensors.** We examine and evaluate the use of optical sensors as part of the attack. Optical sensors are used to measure the light levels and can be sampled at very high rates, hence allowing reception of data at a higher bit rate than standard cameras. We discuss the characteristics of optical sensors and the appropriate modulation method for this type of receiver. We also evaluate various types of cameras, including remote cameras, 'extreme' cameras, security cameras, smartphone cameras, and drone cameras.

The remainder of the paper is organized as follows. In Section II we present related work. Section III describes the adversarial attack model. Technical background is provided in Section IV. Data transmission is discussed in Section V and the implementation in Section VI. Section VII presents the evaluation and results. Countermeasures are discussed in Section VIII, and we present our conclusions in Section IX.

## II. Related Work

The general topic of covert channels used by malware has been extensively studied for more than twenty years. In order to evade the detection of firewalls and IDS and IPS systems, attackers may hide the leaked data within legitimate Internet traffic. Over the years many protocols have been investigated in the context of covet channels, including TCP/IP, HTTPS, SMTP, VOiP, DNS requests, and more [22]. Other types of covert channels include the timing channel in which the attacker encodes the data with packet timing [23], and image steganography [24] in which the attacker embeds the data into an existing image. A sub-topic of covert channels focuses on the covert leakage of data from air-gapped computers, where Internet connectivity is not available to the attacker. Air-gap covert channels, which can be categorized as electromagnetic, acoustic, thermal, and optical channels, have been the subject of academic research for the past twenty years.

### A. *Electromagnetic, acoustic and thermal*

In electromagnetic covert channels, the electromagnetic emission generated by various hardware components within the computer is used to carry the leaked information. Kuhn and Anderson [12] presented an attack ('soft-tempest') involving hidden data transmission using electromagnetic waves emanating from a video cable. The emission is produced when crafted images are transmitted to the screen. In 2014, Guri et al introduced AirHopper [11] [25], a type of malware aimed bridging the air-gap between computers and a nearby mobile phone by exploiting FM radio signals emanating from the video card. In 2015, Guri et al presented GSMem [15], a malware that can generate electromagnetic emission at cellular frequencies (GSM, UMTS, and LTE) from the memory bus of a PC. This study showed that data modulated over the emission can be picked by a rootkit residing in the baseband firmware of a nearby mobile phone. USBee [26], presented in 2016 by Guri et al, used the USB data bus to generate electromagnetic signals and modulate digital data over these signals. Similarly, Funtenna [27] utilized the general GPIO buses of embedded devices to generate electromagnetic signals. Other types of magnetic covert channels are discussed in [28] and [29].

[1] In this paper we refer to networking equipment (LAN switches and routers) as routers.

| Method | Bitrate | Type |
|---|---|---|
| AirHopper [11] [25] (graphic card, video cable)<br>GSMem [26] (RAM-CPU bus)<br>USBee [27] (USB bus)<br>Funthenna [13] (GPIO) | 480 bit/sec<br>1 to 1000 bit/sec<br>4800 bit/sec<br>N/A | Electromagnetic |
| [17] [16] [30] [32] [39] (speakers)<br>Fan noise (Fansmitter) [33] (computer fans)<br>Hard disk noise (DiskFiltration) [32] (Hard-Disk Drive) | <100 bit/sec<br>0.25 bit/sec<br>3 bit/sec | Acoustic |
| BitWhisper [35] (CPU, PC heat sensors) | 1-8 bit/h | Thermal |
| Hard Drive LED (LED-it-GO) [19]<br>Visisploit (invisible pixels)<br>Keyboard LEDs [20]<br>Screen LEDs [21]<br>Implanted infrared LEDs [36]<br>Switches/Routers (the current research) | 15 - 4000 bit/sec<br>< 100 bit/sec<br>150 bit/sec<br>20 bit/sec<br>15 bit/sec<br>10 - 1000 bit/sec (per LED) | Optical |

**Table 1. Different types of air-gap covert channels and their bitrate**

Hanspach introduced a method called acoustical mesh networks in air, which enables the transmission of data via high frequency sound waves [16]. They used laptop speakers and microphones for the transmission and reception. The concept of communicating over inaudible sounds has been studied in [30] and was also extended for laptops and mobile phones. In 2016, Guri et al presented Fansmitter [31] and DiskFiltration [32], two methods enabling exfiltration of data via sound waves, even when the computers are not equipped with speakers or audio hardware. This research show how to utilize computer fans and hard disk drive actuator arms to generate covert sound signals.

BitWhisper, presented in 2015, exploits the computers' heat emissions and the PC thermal sensors to create a so-called thermal covert channel [33]. This method enabled bidirectional covert communication between two adjacent air-gapped computers.

### B. *Optical*

In 2002, Loughry and Umphress discussed the threat of information leakage from optical emanations [20]. In particular, they showed that LED status indicators on various communication equipment carries a modulated optical signal correlated with information being processed by the device. In Appendix A they presented the threat of using the keyboard LED for data exfiltration and were able to achieve a transmission bit rate of 150 bit/sec for each LED with an unmodified keyboard. In the same way, Sepetnitsky and Guri presented the risks of intentional information leakage through signals sent from the screen power LED [21]. The main drawback of these methods is that they are not completely covert: given that the keyboard and screen LEDs don't typically blink, it is possible for users to detect the communication. Recently, Lopes presented a hardware based approach for leaking data using infrared LEDs maliciously installed on a storage device [34]. By blinking the infrared LEDs, malware can leak sensitive data stored on the device at a speed of 15 bit/sec. Note that their approach requires the attacker to deploy the compromised hardware in the organization. In 2017, Guri et al presented a method codenamed LED-it-GO [19], which enables data leakage from air-gapped networks via the hard drive indicator LED which exists in almost any PC, server, and laptop today. They showed that a malware can indirectly control the hard drive LED at a rate of 5800 Hz which exceeds the visual perception capabilities of humans. VisiSploit [35] is an optical covert channel in which data is leaked through a hidden image projected on an LCD screen. With this method, the 'invisible' QR code that is embedded on the computer screen is obtained by a remote camera and is then reconstructed using basic image processing operations. Brasspup [36] demonstrated how to project invisible images on a modified LCD screen. His method is less practical in a real attack model, since it requires physically removing the polarization filter of the targeted LCD screen.

Table 1 summarizes the different types of existing air-gap covert channels and presents their maximum bandwidth and effective distance. Unlike previous work, this study focuses entirely on the network equipment LEDs (switches and routers) as a leaking medium, a threat that has not been studied before. We examine the boundaries of this technique on different types of network equipment and evaluate it with various types of cameras and optical sensors as receivers. With one LED we achieved an exfiltration speed of more than 1000 bit/sec per LED.

## III. Adversarial Attack Model

As a typical covert channel our adversarial attack model consists of a transmitter and a receiver. The transmitter in our case is a network switch or router in which the data is exfiltrated via its LEDs. The receiver is a remote camera or optical sensor which record the LEDs signals. Following we briefly discuss the relevant consideration in a context of the attack model.

### A. *The transmitter*

As part of the attack model the attacker must execute malicious code within the targeted router to enable control of the status LEDs. We present two types of attacks that can be used for this purpose: (1) modifying the firmware on a router, and (2) executing a malicious script or shellcode on an unmodified router.

surveillance, closed-circuit TV, or IP camera positioned in a location where it has a line of sight with the transmitting computer [45], (4) a malicious insider, also known as the evil maid [46], carrying a smartphone or wearable video camera (e.g., hidden camera [47] ) that can position him/herself so as to have a line of sight with the transmitting router, (5) an optical sensor capable of sensing the light emitted from the router LEDs. Such sensors are used extensively in VLC (visible light communication) and LED to LED communication [48]. Notably, optical sensors are capable of sampling LED signals at high rates, enabling data reception at a higher bandwidth than a typical video camera. An example of the air-gap covert channel is provided in Figure 1 in which data is encoded in binary form and covertly transmitted over a stream of LED signals. A hidden video camera films the activity in the room, including the router and LAN switch LEDs. The attacker can then decode the signals and reconstruct the modulated data.

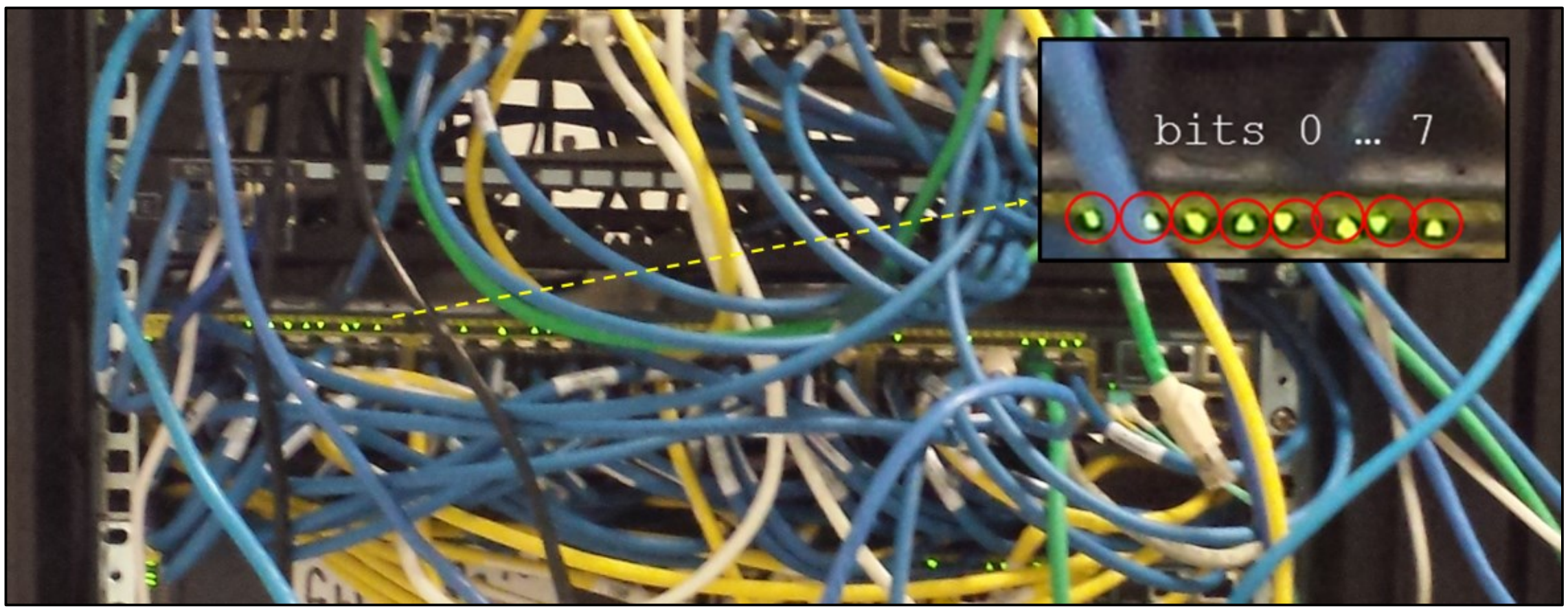

**Figure 1. The attack illustration. Data is encoded in binary form and covertly transmitted over a stream of LED signals emitted from a router or LAN switch.**

**Firmware modification.** In this type of attack the adversary has to infect a router with a malicious firmware. This firmware contains additional code to control the LEDs and encoding a data over it. The Infection of a router can be achieved via supply chain attacks, social engineering techniques, or the use of hardware with preinstalled malware [37] [38] [39]. Notably, in recent years there have been several cases in which routers have been infected with malicious code as a part of an attack. In 2014, *the Guardian* published an article that stated that some network devices were infected by a backdoor before they were delivered to the customer [40].

**Remote code execution.** In this case, the targeted router doesn't have to be infected with a malicious firmware. Instead, it is controlled remotely (e.g., from a compromised computer within the network) via standard remote management channels such as SSH and telnet or by exploiting certain vulnerabilities in the router. The transmitting code is then uploaded to the router in the form of a shellcode or a shell script. Hijacking routers remotely is a common type of attack that has been demonstrated many times in recent years [38] [41]. Other types of flaws and remote code execution vulnerabilities in routers have been found in the wild [42] [43]. For example, in 2017, Cisco found a new critical zero-day vulnerability that affects more than 300 of its switch models. The vulnerability on the Cisco IOS and Cisco IOS XE software, allows remote attackers to remotely execute malicious code on the device and gain root privileges to take full control of the device [44].

### *B. The receiver*

The receiver is a digital camera or optical sensor which has a line of sight with the compromised router's LED panel. There are several types of equipment that can play the role of the receiver in this attack model: (1) a hidden camera that has a line of sight to the transmitting router, (2) a high resolution camera which is located outside the building but positioned so it has a line of sight with the transmitting router, (3) a video

## IV. Technical Background

There are different types of network devices and LEDs. A typical network switch includes two LEDs for each LAN port: the link LED and the status/mode LED. The link LED usually indicates that the port is currently enabled and receiving data from a connected network device. The status/mode LED displays various information about the connection such as mode (half/duplex) and speed (100Mb, 1 GB, etc.). In many types of routers and network switches there are additional custom LEDs for internal fault alerts, fan speed monitoring, and test displays [49]. The most prevalent LED colors in network devices are green and orange, but other colors also in use [49].

### *A. Hardware Level*

At the hardware level, typical network switches and routers consist of an embedded computer with several network controller interfaces. The network interfaces are connected to the main computation unit (via a system bus), in which the traffic is routed and processed. Switches mainly operate in the frames of the data link layer (layer 2), and hence consist of less powerful hardware (CPU, RAM, etc.), than those of routers, which operate on packets in the network layer (layer 3). The device LEDs and button are controlled via GPIO pins connected to the device's PCB (printed circuit board).

The circuit in Figure 2 is a common driver circuit used for router LEDs. The GPIO ports in the microcontroller are based on open collector BJT (Bipolar Junction Transistor) transistor. The LEDs are connected to power supply throw the $R_1$ resistor, which limit its maximum forward current. The R and $R_2$ resistors functions as voltage dividers and provide the voltage DC working point of the LEDs. The LEDs' blinking speed can reach couple of MHz. The LED's maximal blinking speed depend on its physical characteristics, the parasitic capacitance, the galvanic connection to the board, and software or hardware limitations at the controller level.

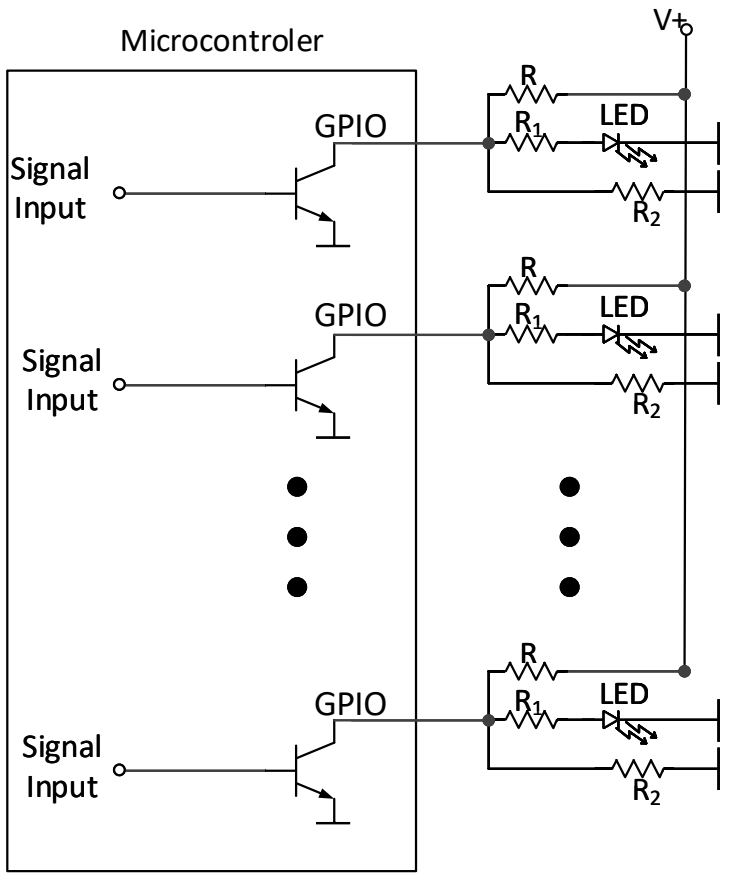

**Figure 2. The hardware microcontroller led driver schematics**

## B. OS Level

There are various proprietary and open-source operating systems for network devices. For example, the Cisco IOS is a Linux based operating system used on most of Cisco's s routers and network switches [50]. The Juniper Network Operating System, Junos OS is a FreeBSD based operating system that is used in Juniper Networks' routers and security devices [51]. Open-source operating systems for network devices include the OpenWrt, DD-WRT, and others [52].

## C. LED Control

In most Linux based OSs, the GPIOs are exported to the user space processes via the GPIO sysfs memory mapped entries. Controlling the status LEDs from a user space is fast enough for most of the applications needs, but still has an overhead of invoking system calls and the context switching from user space to kernel space. When faster access is required, the corresponding GPIOs can be controlled from a kernel driver by directly accessing the I/O pins [53]. Note that sometimes the status LED pins are multiplexed such as they are directly controlled by the hardware. In these cases the corresponding GPIOs must be demultiplexed (via a kernel driver) before the can be controlled by a software.

# V. DATA TRANSMISSION

In this section we discuss the data transmission and describe various modulation methods, along with their implementation details. Note that the topic of visible light communication has been widely studied in the last decade. In particular, various modulations and encoding schemes have been proposed for LED to LED communication [48] [54] [55]. For our purposes, we present basic modulation schemes and describe their characteristics and relevancy to the attack model. As is typical in LED to LED communication, the carrier is the state of the LED, and the basic signal is generated by turning the router LEDs on and off. For convenience, we denote the two states of an LED (on and off) as LED-ON and LED-OFF, respectively.

## A. Data modulation

We present four basic modulation schemes: (1) on-off keying (OOK) – with a single LED, (2) Binary Frequency-Shift Keying (B-FSK) – with a single LED, and (3) Manchester Encoding – with a single LED, and (4) ]Amplitude Shift Keying (ASK) – with multiple LEDs.

### 1) On-Off Keying (OOK) - single LED

On-Off keying is the simplest form of the more general amplitude-shift keying (ASK) modulation. The absence of a signal for a certain duration encodes a logical zero ('0'), while its presence for the same duration encodes a logical one ('1'). In our case, LED-OFF for duration of $T_{off}$ encodes '0' and LED-ON for a duration $T_{on}$ encodes '1.' Note that in the simple case $T_{on} = T_{off}$. In its basic form, this scheme uses a single LED to modulate data. The OOK encoding is described in Table 2.

**Table 2. On-Off-Keying (OOK) modulation**

| Logical bit | Duration | LED state |
|---|---|---|
| 0 | $T_{off}$ | LED-OFF |
| 1 | $T_{on}$ | LED-ON |

### 2) Binary Frequency-Shift Keying (B-FSK) - single LED

Frequency-shift keying (FSK) is a modulation scheme in which digital information is modulated through frequency changes in a carrier signal. In binary frequency-shift keying (B-FSK) only two frequencies, usually representing zero and one, are used for the modulation. In our case, LED-ON for duration of $T_{off}$ encodes logical '0' and LED-ON for a duration $T_{on}$ encodes logical '1.' Note that in the simple case $T_{on} = T_{off}$. We separate between two sequential bits by setting the LED in the OFF state for time interval $T_d$. In its basic form, this scheme uses a single LED to modulate data. The B-FSK encoding is described in Table 3.

**Table 3. Binary Frequency-Shift Keying (B-FSK) modulation**

| Logical bit | Duration | LED state |
|---|---|---|
| 0 | $T_{off}$ | LED-ON |
| 1 | $T_{on}$ | LED-ON |
| Interval | $T_d$ | LED-OFF |

### 3) Manchester Encoding - Single LED

In Manchester encoding each logical bit is modulated using a transition in the physical signal. The sequence of physical signals '01' (LED-OFF, LED-ON) encodes a logical '0' and the sequence of physical signals '10' (LED-ON, LED-OFF) encodes a logical '1.' Manchester encoding's transfer rate is half that of OOK, since it uses two physical signals for each logical bit. This type of encoding is considered more reliable because of the redundancy of each transmitted bit and it is heavily used in communication. The Manchester encoding is described in Table 4.

**Table 4. Manchester encoding, single LED**

| Logical bit | Duration | LED states |
|---|---|---|
| 0 | $T_{on}$ $T_{off}$ | LED-ON, LED-OFF |
| 1 | $T_{off}$ $T_{on}$ | LED-OFF, LED-ON |

*4) Amplitude Shift Keying (ASK) - Multiple LEDs*

In this scheme we use several LEDs to represent a series of bits. As in the OOK encoding, the absence of a signal for a certain time duration encodes a logical zero ('0') for a specific LED, while its presence for the same time duration encodes a logical one ('1') for a specific LED. In the case of multiple LEDs, all LEDs remain in the same status for a duration of $T_{all}$ and then change to the next state. This encoding is relevant for cases where several LEDs in the router are available for the transmission (e.g., a case in which some LAN ports are available). We separate between two sequence of bits by setting the all LED in the OFF state for time interval $T_d$. The ASK encoding for a case of eight LEDs is described in Table 5.

**Table 5. Amplitude Shift Keying (ASK), multiple LEDs**

| Logical bits | Duration | OS operation ($n$=8) |
|---|---|---|
| (0x00) | $T_{all}$ | $LED_1$ = OFF, $LED_2$= OFF, …, LED$n$ = OFF |
| (0x01) | $T_{all}$ | $LED_1$ = ON, $LED_2$ = OFF, …, LED$n$ = OFF |
| (0x02) | $T_{all}$ | $LED_1$ = OFF, $LED_2$ = ON, …, LED$n$ = OFF |
| … | $T_{all}$ | … |
| (0xFF) | $T_{all}$ | $LED_1$ = ON, $LED_2$ = ON, …, LED$n$ = ON |
| Interval | $T_d$ | $LED_1$= OFF, $LED_2$ = OFF, …, LED$n$ = OFF |

## B. Bit Framing

The data is transmitted in a small packets called frames. Each frame is composed of a preamble, a payload, and a checksum (Table 6).

**Table 6. Bit framing**

| Preamble | Payload | Checksum |
|---|---|---|
| 8 bit | 256 bit | 16 bit |

The preamble consists of a sequence of eight alternating bits ('10101010') and is used by the receiver to periodically determine the properties of the channel, such as $T_{on}$ and $T_{off}$ (or the $T_{all}$ in case of multiple LEDs). In addition, the preamble header allows the receiver to identify the beginning of a transmission and calibrate other parameters, such as the intensity and color of the transmitting LED. The payload is the raw data to be transmitted. In our case, we arbitrarily choose 256 bits as the payload size. For error detection, we add a CRC (cyclic redundancy check) value, which is calculated on the payload and added to the end of the frame. The receiver calculates the CRC for the received payload, and if it differs from the received CRC, an error is detected. More efficient bit framing may employ variable length frames, error correction codes, and compression, which are eliminated from our discussion for simplicity.

# VI. IMPLEMENTATION

To evaluate the covert channel we implement a transmitter for the OpenWrt operating system [56]. The OpenWrt is an open-source Linux based OS, used on embedded devices such as routers, gateways, and handheld devices. It supports many type of routers from a wide range of vendors including Cisco, Linksys, D-Link, and others [57]. Using OpenWrt, we were able to develop the LED control module and test it on different types of hardware. Note that the OpenWrt source-code includes many extra features such as advanced routing, firewalling, tunneling, and load balancing.

## A. LED Control

In Linux, the hardware LEDs may be accessed from the kernel space driver or user space process [58]. The kernel space driver can directly access the appropriate GPIO pins in order to turn the LEDs on and off. Such low-level implementation is device specific and requires compilation of the LED driver for the specific target hardware. The LEDs are also exposed to user space processes through the `/sys/class/leds` entries. The entry `/sys/class` is exported by the kernel at run time, exposing the hierarchy of the system abstraction, in order to keep it generic enough to be tested on different hardware. `/sys/class/leds/` contains the properties of each LED, such as name and brightness level. Note that most LEDs don't have hardware brightness support, and hence the brightness value represents only two states (ON / OFF). In addition to the `/sys/class/leds` directory, the kernel may also expose the GPIO interface to the user space through sysfs via the `/sys/class/gpio/`directory [59].

**Algorithm 1** ModulateOOK

```
procedure ModulateOOK(nLED, data, T)
openLED(nLED);            //opens the LED file for writing
while(data[i] !=0)
    if(data[i] == '0')        //modulate 0 by turning the LED off
        setLEDOff(nLED);
    if(data[i] == '1')        //modulate 1 by turning the LED on
        setLEDOn(nLED);
    i++;
    sleep(T);                 // sleep for time period of T
closeLED(nLED);           // closes the LED file descriptor
```

Algorithm 1 shows the pseudocode for the simplest case of OOK modulation. The ModulateOOK procedure receives the target LED number (nLED) to modulate the data on, an array of data to transmit, and the bit duration (T). The LED's brightness file is opened in line 1. The algorithm iterates on the data array and extracts the current bit. If the bit to transmit is '0' then the LED is turned off for time period T (line 5). If the bit to transmit is '1' the LED is turned on for time period T (line 7). We present the implementation only for the OOK modulation; Manchester encoding, FSK, and the ASK with multiple LEDs are omitted for simplicity.

### *B. Shellscript*

We also implemented a version of the transmitter which requires no persistency within the router firmware. In this version, a compromised computer within the network is connecting to the target router though a standard remote connection such as telnet or SSH, or by exploiting a vulnerability from the network. The attacker then executes the transmitting shellcode or script that controls the router LEDs. Note that this type of malware is not persistent and will not survive a router reset. We used a router with a standard DD-WRT firmware that has a telnet server. After connecting to the router from a computer in the network, we execute a script which controls the LEDs and modulates the data. The basic LED control commands used by our script are shown below.

```
// Method #1
// turn the LED on
1: echo 0 > /sys/class/leds/led_name/brightness
// turn the LED off
2: echo 255 > /sys/class/leds/led_name/brightness
// Method #2
3: echo 1 > /proc/gpio/X_out // turn the LED on
4: echo 0 > /proc/gpio/X_out // turn LED off
```

Note that we used two different methods to control the LEDs. In the first method the LED directory `/sys/class/leds/` exported by the kernel is used, and in the second method the GPIO interface `/proc/gpio/` is used. The names of the directory and files may vary between different devices and OSs. Moreover, a router may not expose the LEDs and GPIO interface to the user level (e.g., the appropriate driver has disabled or removed). In this case, the attacker has to execute a kernel level shellcode which directly controls the GPIOs.

## VII. EVALUATION

In this section we evaluate the optical covert channel. Our evaluation focuses on the optical characteristics of router LEDs and the transmission rate. In our experiments we adapt the approach commonly used in visible light communication (VLC), which assumes a line of sight between the light source and the camera [20] [54]. We implemented a prototype of the previously described transmitter on three types of routers that are shown in Table 7. Note that all of them are Wi-Fi routers, and hence are not the typical devices installed on air-gapped networks. However, our goal is to evaluate the characteristics of router LEDs as a medium for a covert channel, so the type of router is less relevant.

**Table 7. The tested routers**

| # | Router | # of LEDs |
|---|---|---|
| R1 | TL-WR841N (TP-LINK) white | 7 + 1 (the power LED) |
| R2 | TL-WR941ND (TP_LINK) yellow | 8 + 1 (the power LED) |
| R3 | AC1750 or Archer c7 (Archer) black | 8 + 1 (the power LED) |

For the evaluation we used the Linux DD-WRT and Linux OpenWrt operating systems. A list of the LED control sysfs entries exposed by the kernel and the on/off values of the LEDs are provided in Table 8.

**Table 8. The tested routers' OS control**

| # | Entry | On/off values | Firmware |
|---|---|---|---|
| R1 | /sys/class/leds/<led_name>/brightness | 0xFF / 0x00 0x00 / 0xFF (depend on the LED) | DD-WRT (version 3.18.48) |
| R2 | /sys/class/leds/generic_<led>/brightness | 0xFF / 0x00 | OpenWrt (Version 3.18.23) |
| R3 | /sys/class/leds/tp-link:green:name/brightness | 0x00 / 0xFF | DD-Wrt (version 3.18.48) |

### *A. Camera receivers*

There are two types of receivers relevant to the attack model: cameras and light sensors. Receiving the optical signals by a camera depends on the line of sight and visibility of the router LEDs. After receiving the recorded video, the attacker has to process the video in order to detect the location of each transmitting LED. The video is processed frame by frame to identify the LED status (on or off) of each frame. Finally, the binary data is decoded based on the encoding scheme. The main factor in determining the maximum bit rate for video cameras is the number of frames per second (FPS). In our experiments, we identified two to three frames per bit as the optimal setting needed to successfully detect the LED transmissions. Table 9 show the maximal bit rate when various types of video cameras are used as receivers.

**Table 9. Maximum bit rate of different receivers**

| Tested Camera/Sensor | FPS | Max bit rate (per LED) | Eight LEDs |
|---|---|---|---|
| Entry-level DSLR (Nikon D7100) | 60 | 15 bit/sec | 120 bit/sec |
| High-end security camera (Sony SNCEB600) | 30 | 15 bit/sec | 120 bit/sec |
| Extreme camera (GoPro Hero5) | 60 - 240 | 100-120 bit/sec | 800-960 bit/sec |
| Webcam (HD) | 30 | 15 bit/sec | 120 bit/sec |
| (Microsoft LifeCam) | | | |
| Smartphone camera (Samsung Galaxy S6) | 30 - 120 | 15-60 bit/sec | 120-480bit/sec |
| Wearable camera (Google Glass Explorer Edition) | 30 | 15 bit/sec | 120 bit/sec |

*B. Light sensor receivers*

A photodiode is a semiconductor that converts light into electrical current. To evaluate the transmissions at high speeds, we built a measurement setup based on photodiode light sensors (Figure 3). The Thorlabs PDA100A light sensor [60] (Figure 3,a) is connected to an internal charge amplifier and a data acquisition system (Figure 3,b). We also used an optical zoom lens to focus the sensing area and reduce the optical noise. The data is sampled with the National Instruments cDAQ portable sensor measurement system [61] via a 16-bit ADC NI-9223 card [62] which is capable of 1M samples per second. The light emitted from the transmitting router (Figure 3,c) is sampled by the sensor and processed by a MATLAB DSP program (Figure 3,d).

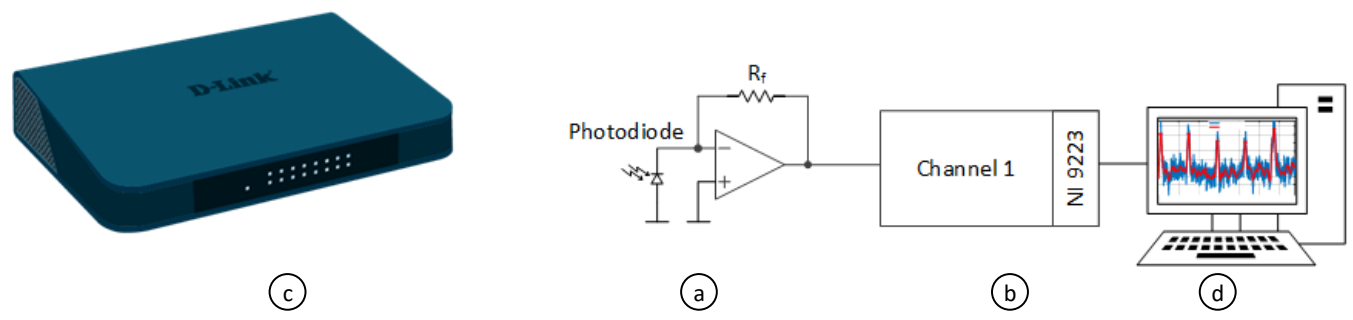

**Figure 3. The measurement setup with the Thorlabs PDA100A light sensor and NI-9233 data acquisition hardware**

*1) Blinking frequency*

In this experiment we tested the maximum frequency at which the router LEDs can blink when controlled from a user space program or shellcode running within the router's OS. The blinking frequency is important, since it defines the maximal speed of the basic signal carried by the LED.

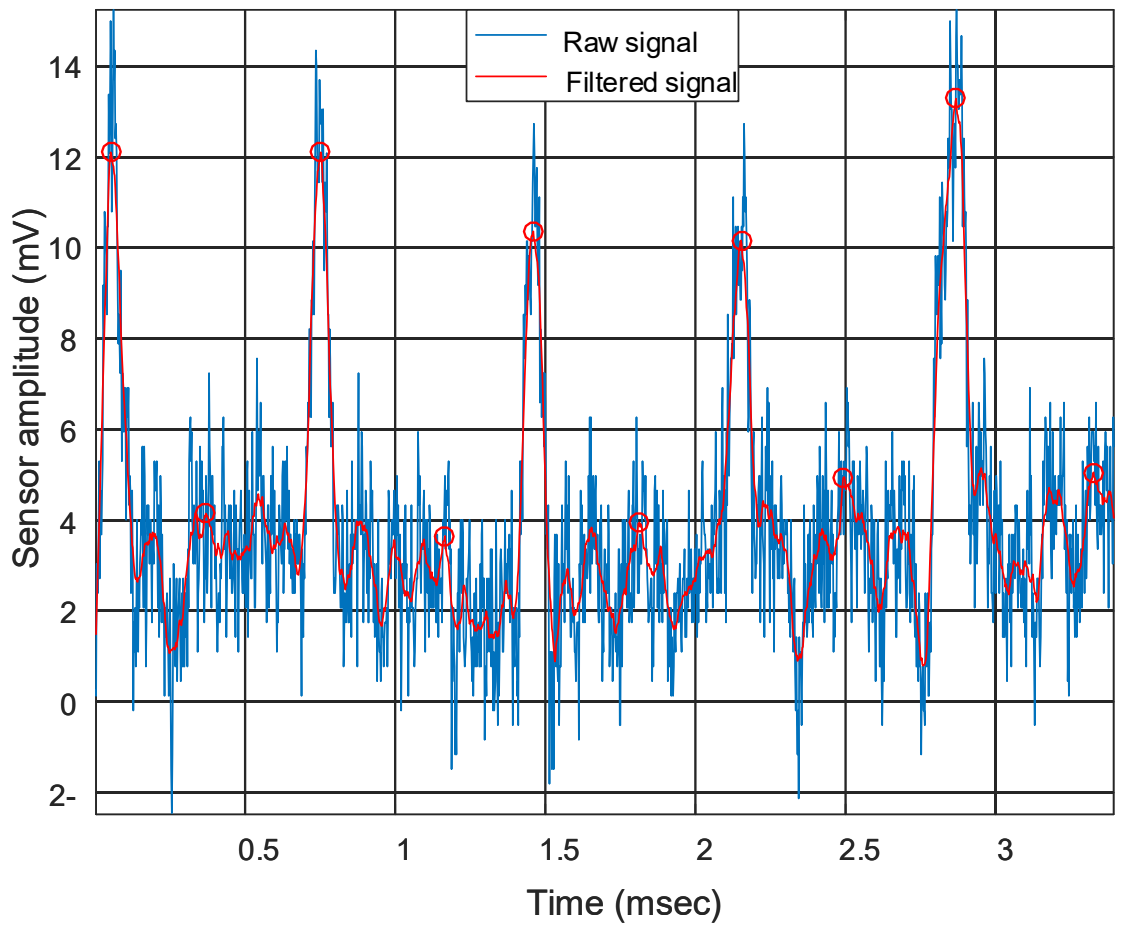

**Figure 4. Maximal blinking frequency (R1)**

Figure 4 shows the signals received from R1 when its leftmost LED is repeatedly turned on and off. The sampling rate in this test is 500K samples per second. As can be seen, the minimal LED-ON time is ~120µs. The minimal blinking time (LED-ON, LED-OFF) is ~700µs which implies a bit rate of 1400 bit/sec in the simplest OOK modulation. During the LED-ON time the sampled amplitude is approximately 14mV, while for LED-OFF the time is 4mV (generated by the ambient light in the room). The blue line is the raw sampled signal, and the red line is the signal smoothed with the Savitzky-Golay filter [63].

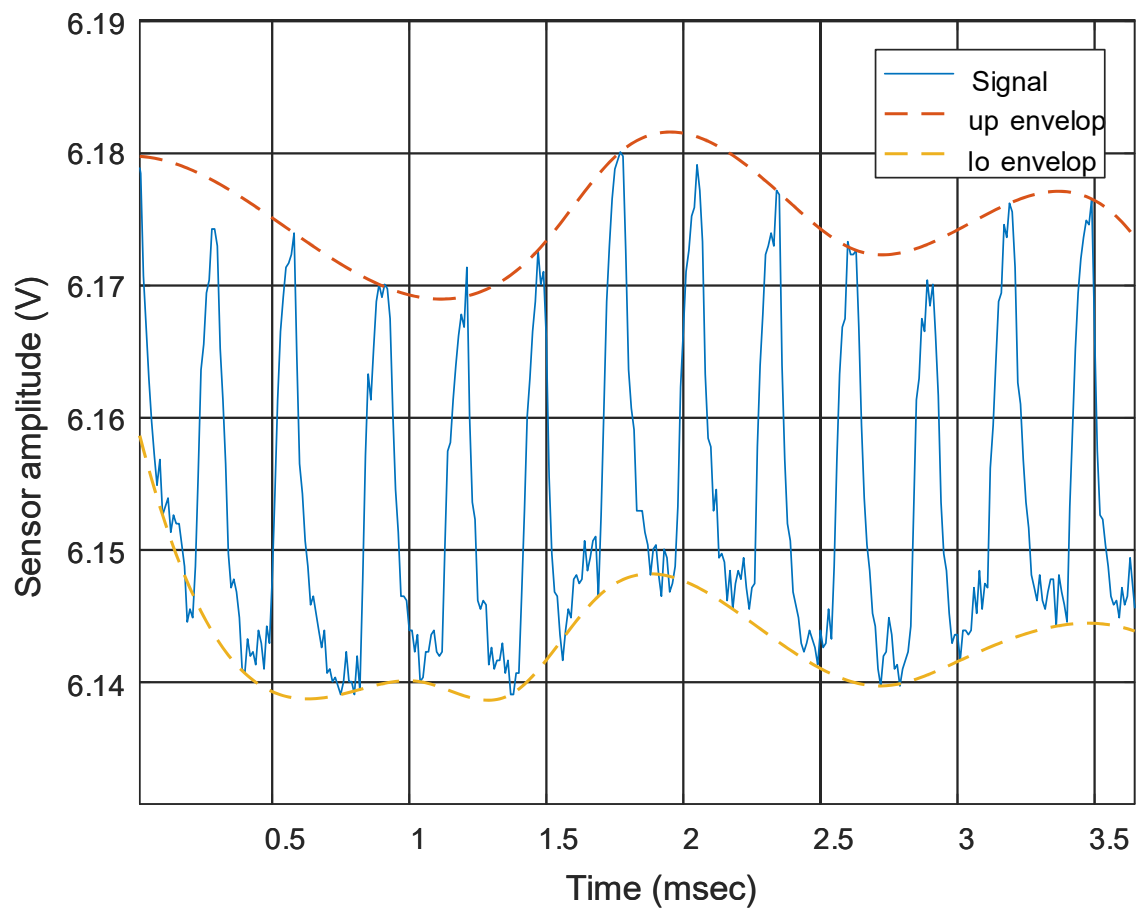

**Figure 5. Maximal blinking frequency (R2)**

Figure 5 shows the signals received from R2 when its leftmost LED is repeatedly turned on and off. As can be seen, the minimal LED-ON time is ~190µs. The minimal blinking time (LED-ON, LED-OFF) is ~290µs which implies a bit rate of 3450 bit/sec with the simplest OOK modulation. During the LED-ON time the sampled amplitude is approximately 30mV, while for LED-OFF the time is 6.14V. As can be seen in Figure 5, the room's backlights are flickering at 500 Hz (the dashed line).

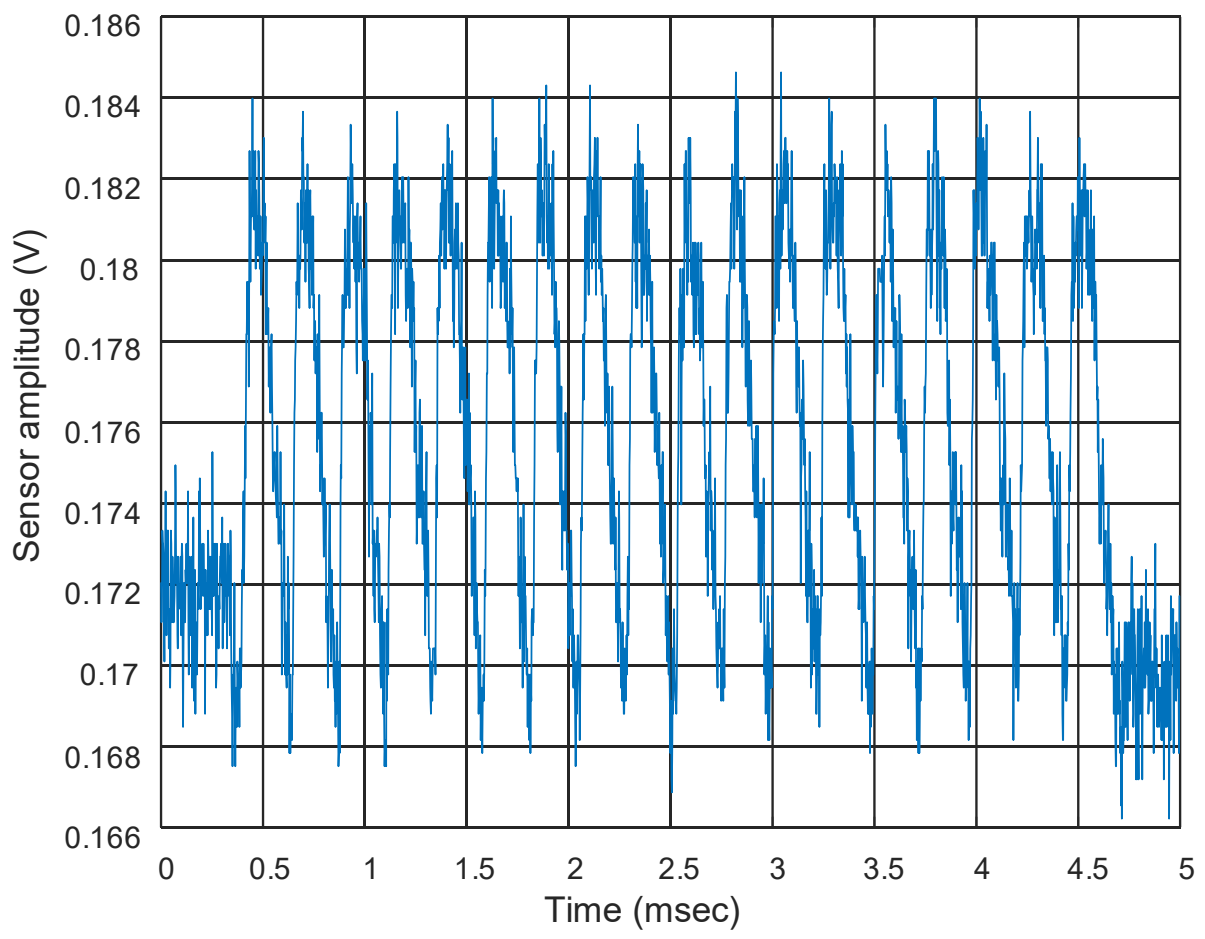

**Figure 6. Seven LEDs blinking at maximal frequency (R1)**

Figure 6 shows the signals received from R1 when all seven LEDs are repeatedly turned on and off. By using all of the LEDs together for modulation, we significantly increase the optical signals emitted from the transmitting router. This method can be used when the optical signal generated by a single LED is too low for successful reception. As can be seen, with multiple LEDs the minimal blinking time (LED-ON, LED-OFF) is ~240μs which implies a bit rate of 4000 bit/sec at the simplest OOK modulation.

### *2) Amplitude Modulation*

With a camera receiver it is possible to distinguish between two or more different transmitting LEDs. In this case the bit rate is derived from the number of LEDs available for modulation. That is, with $N$ LEDs we can generate $2^N$ different signals. Unlike camera receivers, light sensors can only measure the *amount* of light emitted from the router and not distinguish between different LEDs. One straightforward strategy is to use OOK modulation when 0 is modulated with all of the LEDs in the OFF state, and 1 is modulated with all of the LEDs in the ON state. Obviously, this type of modulation limits the transmission rate. We found that under some circumstances it is also possible to distinguish between different amounts of light emitted when using different number of LEDs, even with a light sensor. Consequentially, we are able to increase the bit rate by modulating multiple bits with several LEDs (using ASK modulation) when a light sensor is used for reception. Under optimal conditions $n$ different amplitudes can modulate $log_2(n)$ values.

Figure 7 and Figure 8 shows eight amplitude levels as measured from R1 when all seven LEDs are in use. We employ eight different states, starting with all seven LEDs in the off state and sequentially turning the LEDs on until all of the LEDs are on (0000000, 1000000, 1100000… 1111111).

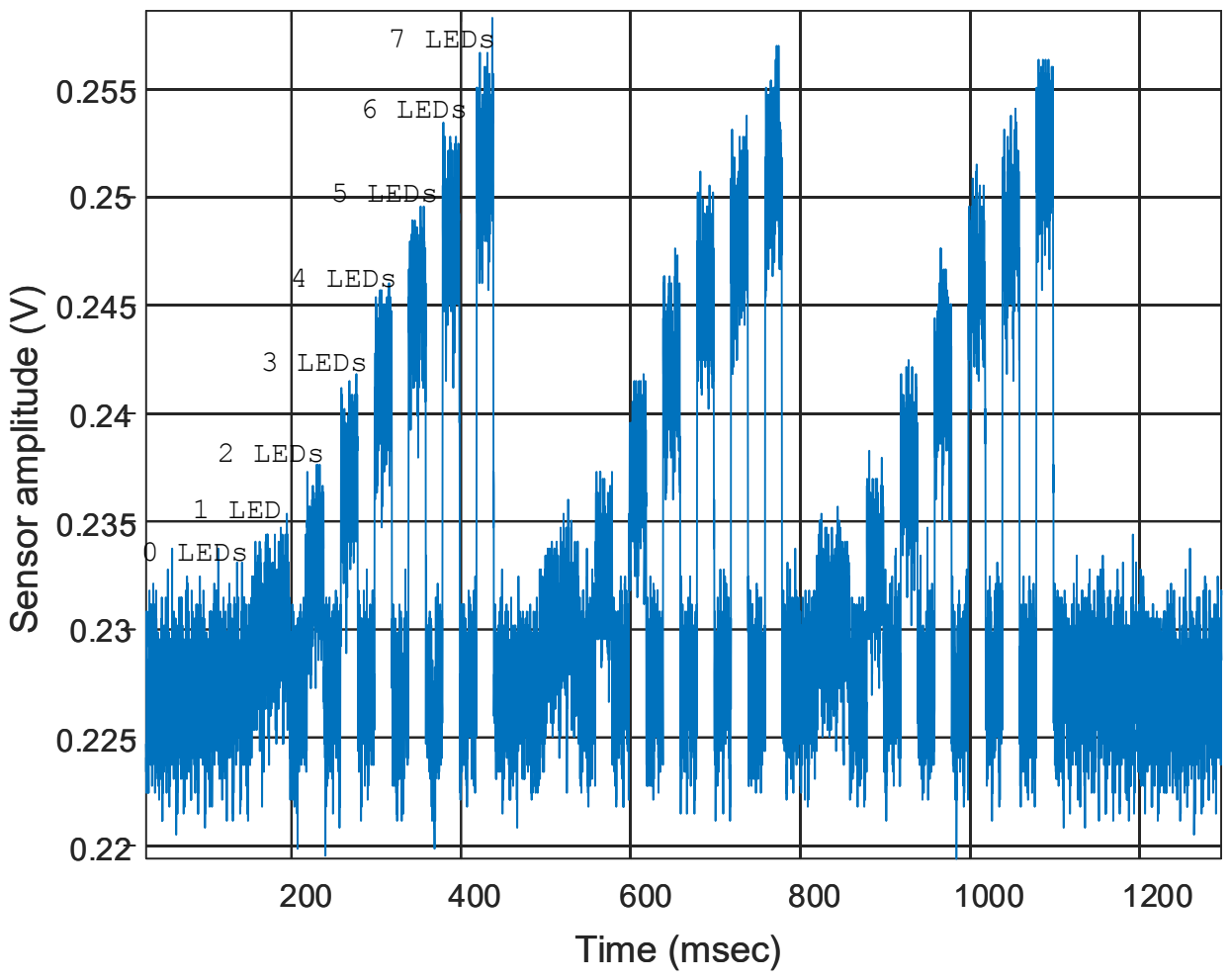

**Figure 7. Eight amplitudes modualted by seven LEDs, 10ms per amplitude (R1)**

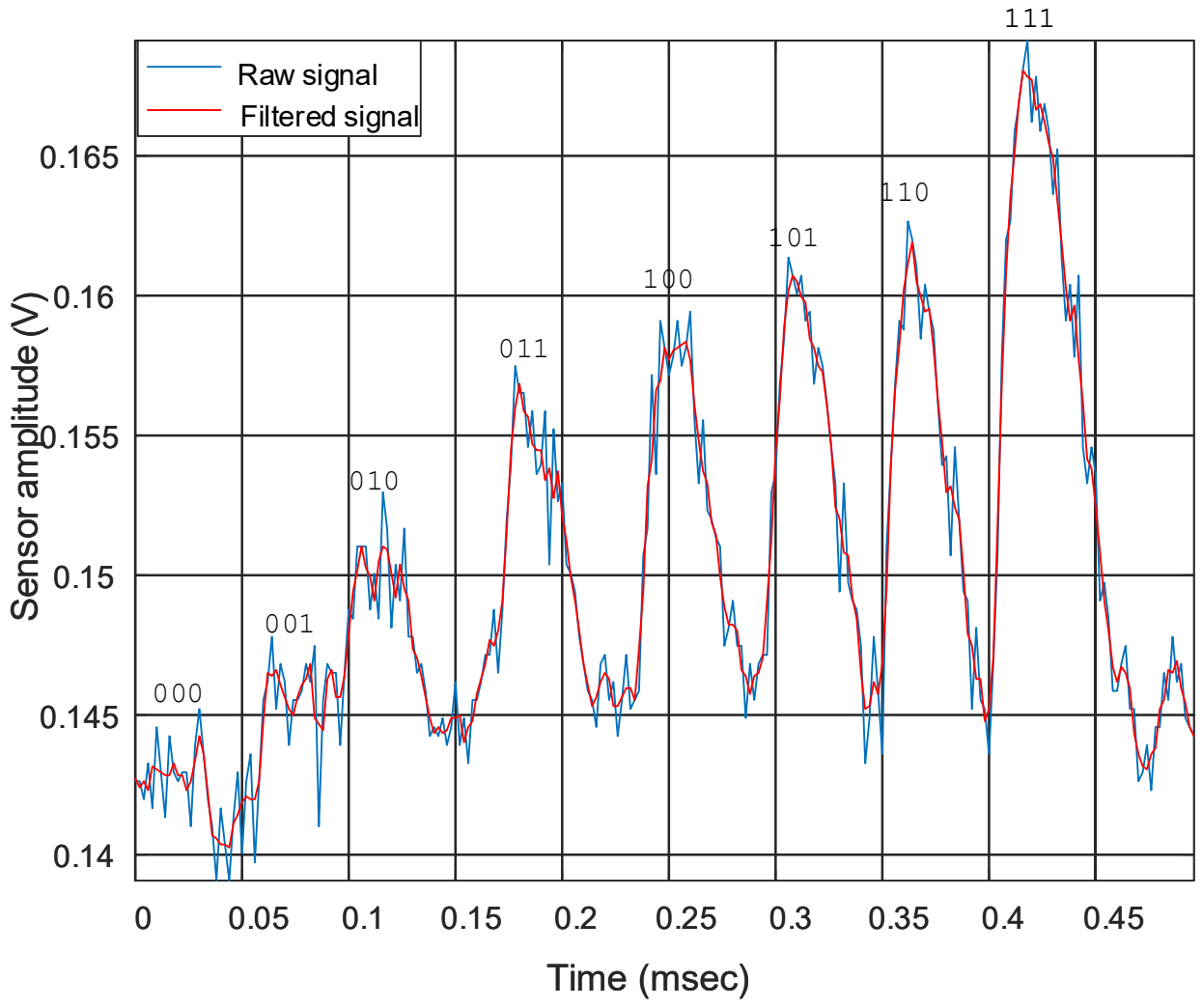

**Figure 8. Eight amplitudes modulated by seven LEDs, 300μs per amplitude (R1)**

Note that we only distinguish between the numbers of LEDs turned on, as opposed to their location (e.g., the states 1100000, 0110000, and 0011000 represent the same amplitude). As can be seen in Figure 8, we can distinguish between eight different levels, when each amplitude level is modulated over 300μs. This implies a rate of ~3.3K levels per second or 10000 bit/sec. In Figure 8 the blue line is the raw sampled signal, and the red line is the signal smoothed with the Savitzky-Golay filter [63].

### *3) Transmission*

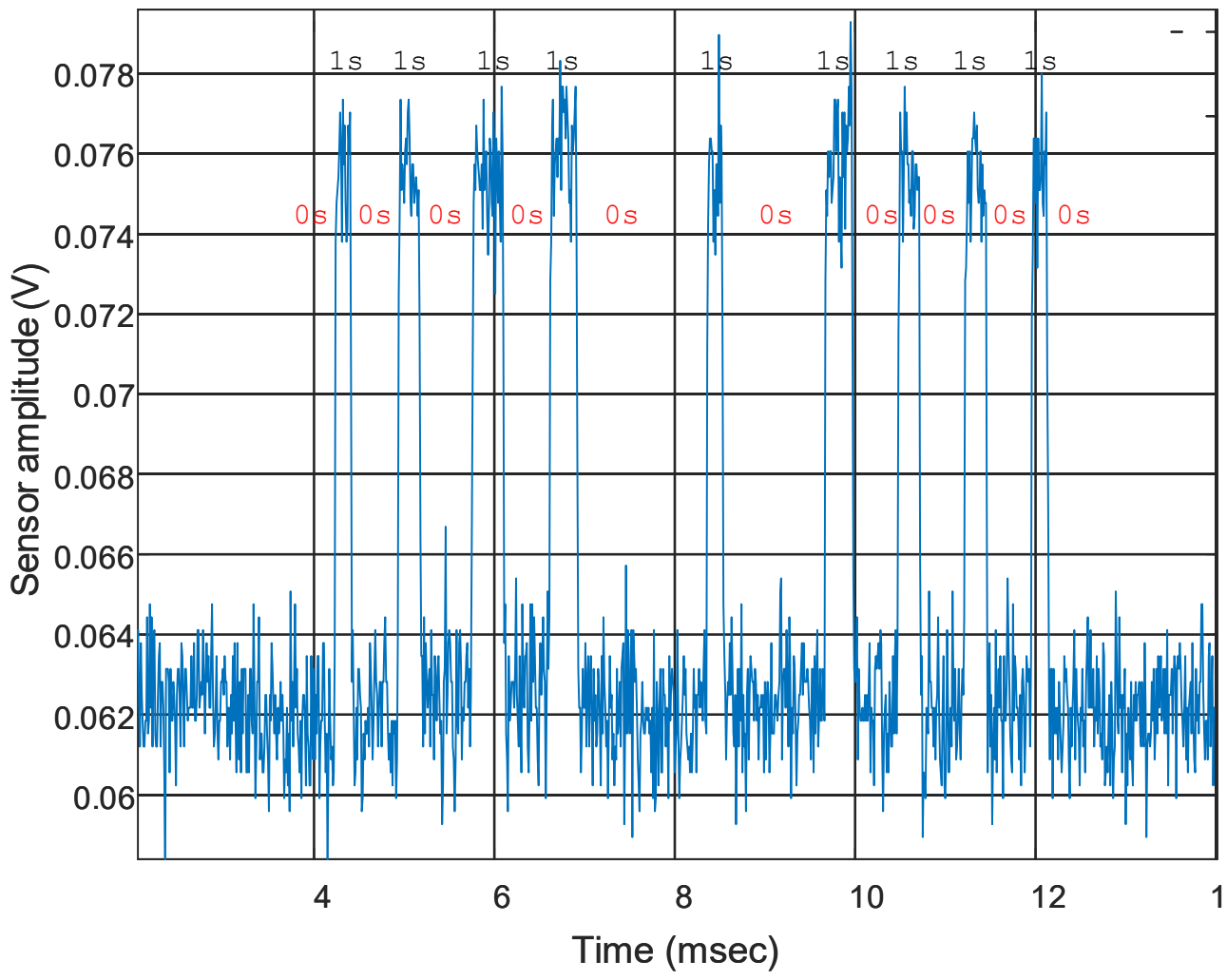

**Figure 9. OOK modulation with one LED (R2)**

Figure 9 shows the measurements in which 32 bits (01011011101110010011101101101010) are transmitted from R2 using the OOK modulation via a single LED. The 32 bits were transmitted in 9ms which implies a bit rate of 3555 bit/sec. The bit error rate (BER) measured in this case was under 5%.

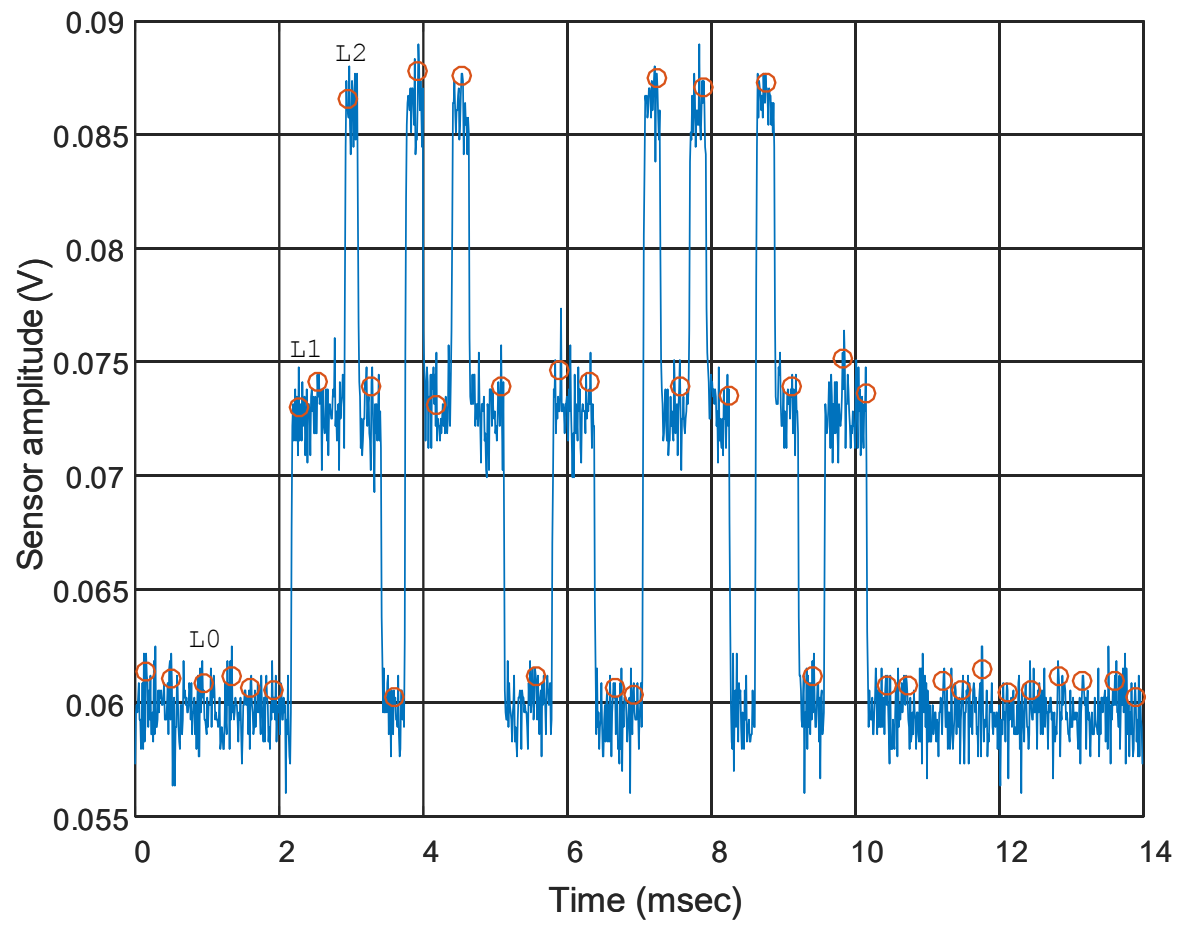

**Figure 10. ASK modulation (R2, two LEDs)**

Figure 10 shows the measurements in which 32 bits (`01011011101110010011101101 1010`) are transmitted from R2 using the ASK modulation via two LEDs. The 32 bits were transmitted in 9ms which implies a bit rate of 3555 bit/sec. Note that the 32 bits are encoded with three amplitude levels (L0, L1, and L2). The BER measured in this case was again under 5%.

## VIII. COUNTERMEASURES

Prevention and detection are the two main types of countermeasures targeted at emanation based data leakage.

**Prevention**. Common countermeasures may include policies aimed to restrict the accessibility of network equipment by placing it in classified rooms where only authorized staff may access it. Typically, all types of cameras are banned from such secured rooms. The NATO TEMPEST standards such as the NATO SDIP-27 (levels A/B/C) and SDIP-28 define classified zones which refer to the perimeter needed to prevent accidental or intentional leakage of signals [64] [65] [66] [67]. In these areas, the presence of surveillance cameras may serve as a deterrence measure. However, as mentioned previously in the attack model, the surveillance camera itself may be compromised by a malware [68] [45]. A less elegant countermeasure against LED attacks is to cover the status LEDs with black tape to physically block the optical emanation [20]. Covering the switch and router LEDs may affect user experience, for example, when testing the device or checking its functionality. Device and cable shielding is another common countermeasure recommended by the NATO standards to address other types of TEMPEST attacks, particularly electromagnetic TEMPTEST attacks. A special window film that prevents optical eavesdropping may be installed [69]; note that this type of countermeasure doesn't protect against insiders attacks, or cameras located within the building. Another approach is to interrupt the emitted signals by intentionally invoking random LED blinking. In this way, the optical signal generated by the malicious code will get mixed up with random noise. Implementing such a noise generator within an embedded device requires the involvement of OEMs and may also affect the usability of the LEDs.

**Detection**. Technological countermeasures may include the detection of the presence of malware that triggers the router status LED. Mitigating firmware level attacks involves detecting (statically or dynamically) whether the firmware of a router has been compromised. Although the detection of firmware attacks has been studied in the recent years [70], most solutions focus on detecting the attack in the network traffic. Detecting an already compromised firmware installed within an embedded device such as a switch or router is still a challenging task [71]. The comparison of the device firmware with a clean image to identify malicious changes has been proposed [72], however extracting the device firmware is not always possible. More recently, Guri et al. proposed using the device's JTAG debugging interface to extract the memory for security purposes [71]. Such a method is considered invasive and involves opening the device for physical forensic investigation, and hence it is not a practical scalable solution. Addressing a software level attack in which the network device is compromised via a computer within the network is possible using conventional intrusion detection and prevention techniques. This type of signature and behavioral based defense has been shown to be limited in detecting zero-day and advanced attacks [73]. Another possible countermeasure is monitoring the router LEDs in order to detect covert signaling patterns. Again, practical implementation would be difficult, because most network device LEDs routinely blink due to frequent traffic activity. Consequentially, this kind of external monitoring solution would likely suffer from a high rate of false alarms.

The countermeasures are summarized in Table 10.

**Table 10. List of countermeasures**

| Type | Countermeasure | Limitations |
|---|---|---|
| Prevention | Zoning, camera banning, and area restriction | Insider threats (e.g., cameras in the building and insiders) |
| Prevention | LED covering | Degradation of functionality and user experience |
| Prevention | Window shielding | High price |
| Prevention | Signal jamming | Requires OEM intervention, Degradation of user experience |
| Detection | LED activity monitoring (external camera) | Price, false alarms |
| Detection | Firmware forensic extraction and investigation (e.g., JTAG). | Technical challenges (e.g., extracting the device image), invasive operation |
| Detection | Malicious traffic detection (software attack) | Limited detection rate (e.g., zero-day and advanced attacks) |

## IX. Conclusion

Network devices (e.g., LAN switches and routers) typically include activity and status LEDs in various sizes and colors. Such LEDs are used for monitoring the traffic activity, providing alerts in cases of hardware or software failure, testing the device, and so on. We show that these LEDs can be controlled by a malicious code which runs on the device. In a typical attack model, data such as files, encryption keys, and keylogging data is encoded and modulated over the LED's blinking patterns. An attacker with a remote camera or optical sensor with a line of sight with the transmitting equipment can receive the data and decode it back to a binary information. We examine the internal architecture of network switches and routers at the hardware and software level and show how their LEDs can be controlled programmatically by accessing the corresponding GPIO controls. We developed a prototype of the transmitter using open-source firmware (OpenWRT and DD-WRT) and tested it on three different home routers. We investigate the covert channel with different types of receivers, including digital cameras and optical sensors. We also discuss different detection and prevention countermeasures. Our research shows that data can be exfiltrated from an air-gapped network via switch and router LEDs at a bit rate of tens to thousands of bits per second per LED, depending on the type of receiver and the number of LEDs in use.